# Elastic constants of single-crystalline NiTi studied by resonant ultrasound spectroscopy

*Lucie Bodnárová[1], Michaela Janovská[1], Martin Ševčík[1], Miroslav Frost[1], Lukáš Kadeřávek[2], Jaromír Kopeček[2], Hanuš Seiner[1], Petr Sedlák[1]*

[1]Institute of Thermomechanics, Czech Academy of Sciences, Dolejškova 1402/5, 18200 Praha 8, Czechia

[2]Institute of Physics, Czech Academy of Sciences, Na Slovance 1999/2, 18200 Praha 8, Czechia

**Abstract:** Contactless, laser-based resonant ultrasound spectroscopy was utilized to monitor changes in elastic properties in single-crystalline NiTi shape memory alloy. It was observed that the elastic behavior of the temperature-induced B19' martensite, which is formed by a fine mixture of variants, adopts the symmetry elements of the parent austenite phase, and thus, the changes over the transformation temperature can be represented by the temperature evolution of three cubic elastic coefficients. The experiments confirm that the transition during the cooling run is preceded by pronounced softening of the $c_{44}$ elastic coefficient, which leads to nearly complete vanishing of elastic anisotropy prior to the transition. Below the transition, this coefficient remains soft, and the character of anisotropy switches from $c_{44}/c'>1$ to $c_{44}/c'<1$. We rationalize this behavior from the mechanical instability of the B19' lattice with respect to shears along the $(001)_{B19'}$ plane, which is known from first-principles calculations.

**Introduction**

NiTi shape memory alloy (SMA) is the most commonly used shape memory material and one of the most studied in the literature. Despite significant progress in describing the properties of NiTi, some fundamental features have not been fully understood yet, and are still the subject of active research. Already decades ago, it was discovered that NiTi alloy behaves very differently compared to other shape memory alloys in several aspects. Very unusual properties were observed in the elasticity and its evolution with temperature. Several studies [1-3] have documented that the austenitic shear elasticity increases with increasing temperature as common for the high-temperature phase in SMA; however, its elastic anisotropy decreases when approaching martensitic transformation (MT) unlike in other SMAs (e.g., Cu-based [4] or NiMnGa [5, 6]) in which instability of austenite before MT is related to strong elastic anisotropy. On the other hand, martensite, whose elasticity is much less documented in the literature, reveals the opposite effect. We can conclude from the strong dependence of the effective elastic properties of martensite on the previous loading history of polycrystalline samples [7-9], the large elastic anisotropy of oriented martensitic microstructures [10, 11], and theoretical calculations based on density functional theory (DFT) [12-15], that martensite exhibits strong elastic anisotropy, and this anisotropy remains large even at low temperatures far below the MT. Although these elastic properties of NiTi have many important consequences, both in the formulation of constitutive models describing its complex behavior [16-17] and in their relation to other unusual properties of NiTi, such as the easier plastic forming of martensite than austenite [18-19], one can find very few experimental works investigating the elastic properties of NiTi for both austenite and martensite.

In this study, we present ultrasonic measurements of the elastic anisotropy and temperature evolution of the elastic constants of single-crystal NiTi austenite and temperature-induced martensite formed by a fine mixture of variants. We use a new experimental approach based on modal resonant ultrasound spectroscopy (RUS) [20,21], allowing us to detect changes in the character of elastic anisotropy and its symmetry. We document the unusual evolution of elasticity in NiTi during MT, which agrees with previous experiments and theoretical studies. Although the effective elasticity of twinned, temperature-induced martensite obtained from experiments cannot completely verify the theoretical DFT predictions of martensite single-variant elasticity, employing the Hill homogenization scheme allows us to discuss the accuracy of the elasticity predictions related to martensite instability with respect to (001)[100]$_{B19'}$ shear, which plays a significant role in the mechanics of martensite.

**Materials and methods**

The experiments were performed on a NiTi single crystal alloy (purchased from Prof. Y. I. Chumlyakov, Tomsk State University) with a nominal composition of 51% Ni and 49% Ti and a 6.5 g/cm$^3$ mass density. For ultrasonic measurements, the material was cut approximately in the principal directions of austenite to obtain rectangular cuboid specimens with dimensions of 3.5 x 3.2 x 2.8 mm$^3$.

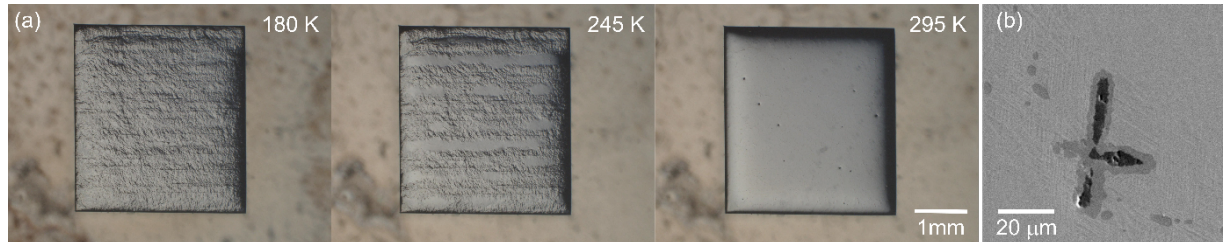

Fig. 1 Optical microscope images of the sample at three temperatures during heating of the sample (a), Image Quality (IQ) snapshot by EBSD method (b) of Ti$_2$Ni precipitates (dark grey) and TiC inclusions (black) within NiTi matrix (light grey).

The microstructure of the samples was probed using a Tescan FERA 3 scanning electron microscope (SEM) (Tescan, Brno, Czech Republic) equipped with an Octane Super energy-dispersive spectroscopy (EDS) analyzer and a DigiView V electron backscatter diffraction (EBSD) method, both from EDAX (Ametek Inc., Berwyn, USA). The data were evaluated with EDAX OIM 8.6 software. In addition to the NiTi matrix, the sample contained a low content (<2%) of finely spread TiC inclusions and Ti$_2$Ni precipitates segregated around the carbide inclusions; see an illustrative snapshot of the microstructure in Fig. 1(b).

The transition temperatures for the examined material were determined by differential scanning calorimetry (DSC) using a TA Discovery DSC 2500 (Waters, New Castle, USA) apparatus; the DSC measurements were made on the sample used for RUS characterization in the temperature range 130 K–370 K with heating and cooling rate of 2 K.min$^{-1}$.

The resonant spectra of the sample were measured using a contactless, laser-based RUS technique [20,21]. The experiments were performed in a temperature-controlled chamber under low pressure nitrogen atmosphere (approximately 20 mbar) in the temperature range 162 K–295 K, so the entire transformation sequence was covered. Vibrations of the sample were excited on the bottom face of the sample by infrared pulse Nd:YAG laser (Quantel ULTRA, nominal wavelength 1.064 μm, pulse duration eight ns) and detected on the upper side of the sample by scanning laser vibrometer (MSA-600) installed in the microscope objective. The resonant spectra were recorded in the frequency range of 0.5–2.5 MHz. The spectra and the modal shapes corresponding to the individual resonant peaks were recorded by scanning laser-Doppler interferometry at room temperature (295 K). The velocities of longitudinal waves in directions perpendicular to individual sample faces were then measured by the pulse-echo method. The same procedure was repeated with the sample cooled in liquid nitrogen to 245 K, i.e., in a fully martensitic state.

To investigate the temperature variation of elastic properties, the evolution of resonant spectra in the temperature range 162 K–295 K was measured in a cryogenic chamber. The temperature stability of the sample was ±1 K. First two resonant frequencies were traced through the whole cycle. The evolution of velocities of longitudinal waves in a direction perpendicular to the largest sample face was measured by the pulse-echo method in the same temperature range.

Figure 1(a) presents snapshots of the sample captured by optical microscope Keyence VHX-7000 with 50x magnification coupled with vacuum chamber Linkam THMS350V with liquid nitrogen cooling. These snapshots revealed that during the temperature-induced MT, the crystal gradually fills with a fine martensitic microstructure. In other words, we did not observe any macroscopic formation of a planar interface between austenite and martensite (habit planes) during the transition, and there were no macroscopically homogeneous regions (simple laminates) visible after the transition had been accomplished. Also, the fine microstructure nucleated and grew on the observed surface in the form of isolated islands embedded in the austenite matrix (often being elongated and coalescing into broader bands), which indicates that it consists of a mixture of different variants without macroscopic shape strains with respect to austenite. We rationalize the observation by the presence of finely dispersed TiC and $Ti_2Ni$ precipitates in the crystal, which precluded the formation of any of the larger-scale features during the transition. The self-accommodated nature of the thermally induced microstructure also suggests that the microstructure does may adopt the symmetry elements from the cubic matrix and, thus, that the transition may not alter the cubic symmetry of the elastic behavior of the crystal. This estimate is confirmed below from the ultrasonic measurements.

**Results and discussion**

Elastic constants of austenite and martensite

First, the RUS measurements of austenite at 295 K and martensite at 245 K were evaluated. The elasticity of single-crystal austenite is cubic with three independent elastic constants: $c_{11}$, $c_{12}$, and $c_{44}$. The high number of measured resonance modes (40 resonant modes were fitted in the inverse procedure) together with the pulse-echo measurement allowed us to determine these three constants with a high accuracy, see Tab. 1. The evaluation of elastic properties of martensite was less straightforward. The temperature-induced martensitic microstructure can be expected to preserve the cubic symmetry of austenite because no particular martensitic variant is preferred during simple cooling. Nevertheless, various factors could violate this assumption (e.g., internal stress in the measured sample).

For this reason, we first evaluated the RUS spectra of martensite without any specific assumption on symmetry—i.e., we allowed for the most general (triclinic) symmetry with 21 independent elastic constants, used the RUS spectrum to determine these 21 constants, and then analyzed the symmetry of the resulting elastic tensor.

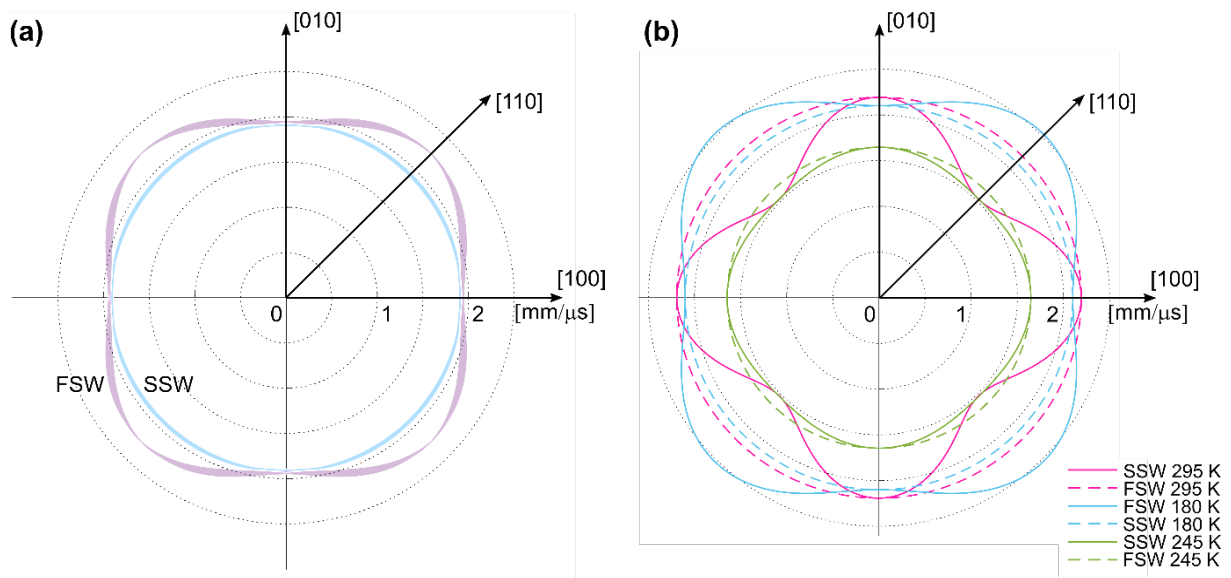

Fig. 2. Distribution of equivalent velocity surfaces of slow (SSW) and fast shear (FSW) waves in the principal $(001)_{B2}$ plane for pure martensite at 160 K (a), and velocity surfaces of SSW and FSW in the same principal plane at three different temperatures, 180 K, 245 K and 295 K, (b). See text for details.

Fig. 2(a) shows an example of such an analysis; it presents the angular dependence of the velocity of propagation of fast and slow shear elastic waves (FSW and SSW, respectively) in martensite in a plane corresponding to the principal $(001)_{B2}$ plane in austenite. For the plot, each direction in this plane represents the set of equivalent crystallographic directions in austenite. For each such set of directions, we calculated the FSW and SSW velocities using the triclinic elasticity tensor obtained from the RUS data. The curve's width depicts the scatter in the resulting set of velocities. The preservation of the cubic symmetry is evident from the very narrow distributions of plotted velocities, which would shrink to a single value in any given direction for a perfect cubic symmetry. The elasticity of martensite can be, thus, assumed as cubic and fully represented by three elastic constants, too. The resulting elastic constants of both phases are summarized in Tab. 1.

Table 1. Elastic constants $c_{11}$, $c_{12}$, $c_{44}$, $c' = (c_{11} - c_{12})/2$ and anisotropy factor $A$ of austenite at 295 K and martensite at 245 K.

|  | $c_{11}$ [GPa] | $c_{12}$ [GPa] | $c_{44}$ [GPa] | $c'$ [GPa] | $A$ [1] |
|---|---|---|---|---|---|
| Austenite (295 K) | 171.5 ± 0.2 | 141.3 ± 0.2 | 31.24 ± 0.05 | 15.09 ± 0.05 | 2.07 |
| Martensite (245 K) | 192.6 ± 0.3 | 126.9 ± 0.3 | 23.94 ± 0.07 | 32.81 ± 0.07 | 0.73 |

Temperature evolution of elasticity during MT

The temperature evolution of all three elastic constants (Fig. 3) was determined from the temperature measurement of resonant spectra and pulse-echo measurement of phase velocities. Figures 3(a) and (b) show the evolution of the constant $c_{11}$ and the two principal shears in cubic symmetry, $c_{44}$ and $c' = (c_{11} - c_{12})/2$. For a direct comparison of the evolution of elasticity with the course of MT, the DSC measurement is plotted in the same figure, see Fig. 3(d), and the corresponding positions of transformation temperatures are indicated. Fig. 4(a) presents the temperature evolution of Zener's anisotropy factor $A$ of the sample, which is the scalar measure of the strength of anisotropy, reaching A = 1 for a perfectly isotropic material; the factor is defined as the ratio between the two shear moduli, $A = c_{44}/c'$, and in its limits, $A \rightarrow \infty$ or $A \rightarrow 0$, it represents the loss of stability of the cubic lattice of austenite with respect to either $[-110](110)_{B2}$ or $[100](010)_{B2}$ shearing, respectively.

The temperature evolution of shear elastic moduli in austenite before MT corresponds well to previous published works [1-3]. The lower shear modulus $c'$ decreases slightly with decreasing temperature. On the other hand, the $c_{44}$ modulus decreases sharply, and its evolution with temperature is strongly non-linear, which leads to the decrease of $A$. Extrapolating these trends into the MT region, we infer that both shear moduli of pure austenite would approach almost the same value during MT, i.e., austenite would become almost elastically isotropic if the transformation to martensite was hindered.

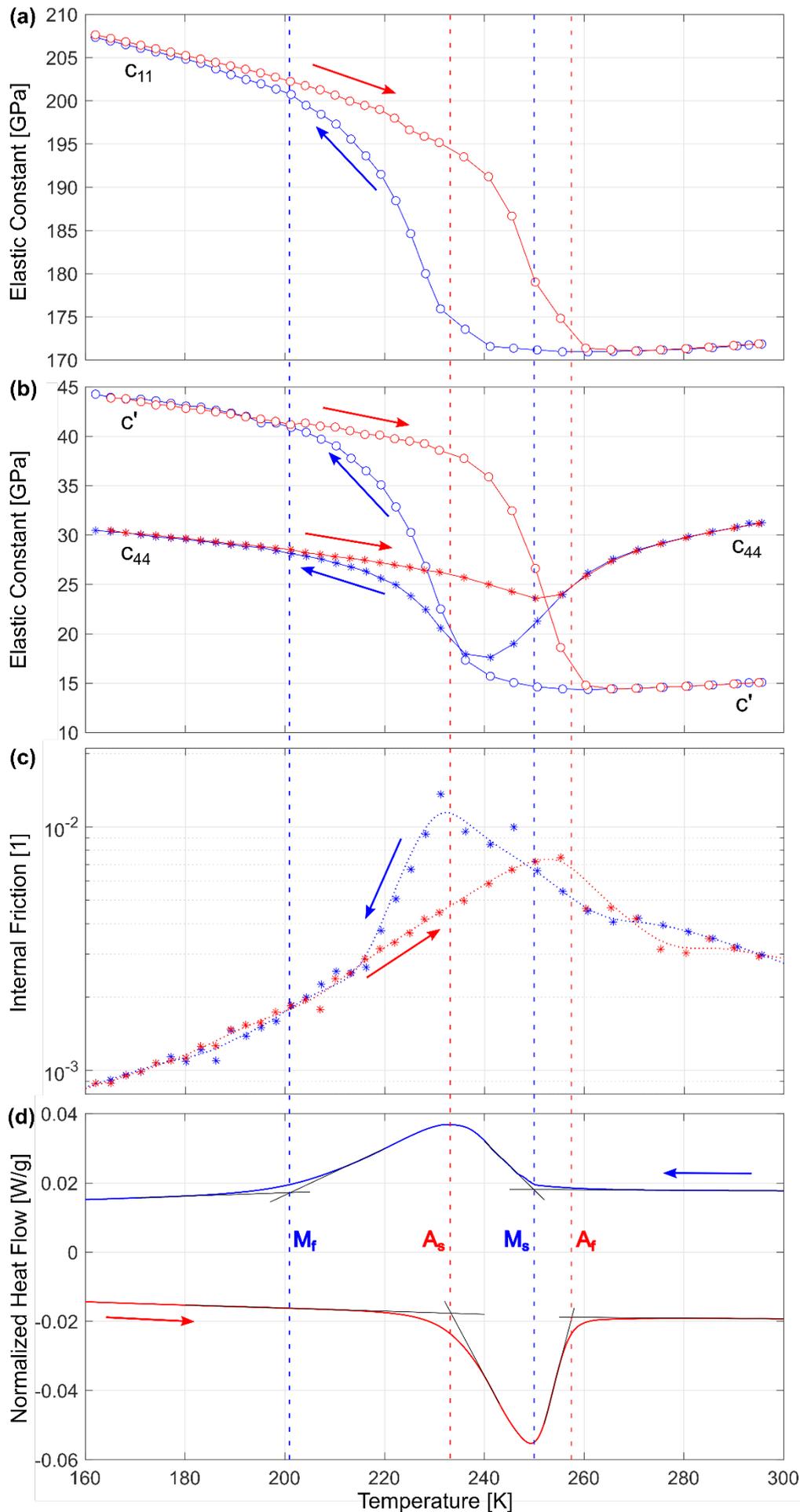

Fig. 3. Temperature dependence of the elastic constants $c_{11}$ (a), $c'$ and $c_{44}$ (b), and of internal friction related to $c_{44}$ (c) within a temperature cycle from 295 K to 162 K (in blue) and back (in red). Corresponding DSC curve (d) with the indicated position of transition temperatures: $M_f$ = 202 K, $A_s$ = 234 K, $M_s$ = 249 K, and $A_f$ = 257 K.

Such a rapid drop in *A* of austenite when approaching the MT temperature from above contrasts with its behavior in most other shape memory alloys, where elastic MT precursors in austenite are usually associated with a sharp increase in elastic anisotropy. The unusual behavior of NiTi is related to the volume-preserving stretching in the $[111]_{B2}$ direction, which is the dominant shear mechanism of B2-B19' transformation and manifests itself dominantly as the plunge of $c_{44}$.

The temperature evolution of elasticity in martensite is less documented in the literature, and for a direct comparison, only the previous results in [2] are available. In accordance with their data and with the assumption of energy stability of the martensitic phase, both shear constants increase with decreasing temperature, see Fig. 3(a) and (b). Interestingly, MT leads to an "anisotropy switch", i.e. the shear modulus $c_{44}$ is lower than $c'$ after the transformation is finished, which results in *A* being close to 0.7 and remaining almost constant during cooling. Similar switching from A>1 to A<1 also occurs in epitaxially grown NiTi films, as reported recently in [22].

Considering that the measured martensite consists of a twinned microstructure with equally populated variants and that the elastic anisotropy of an individual martensite variant is usually weaker than the anisotropy of austenite in SMA [23], the observed difference in $c_{44}$ and $c'$ constants is surprisingly large. The origin of this large anisotropy will be clarified in the next section utilizing the DFT results.

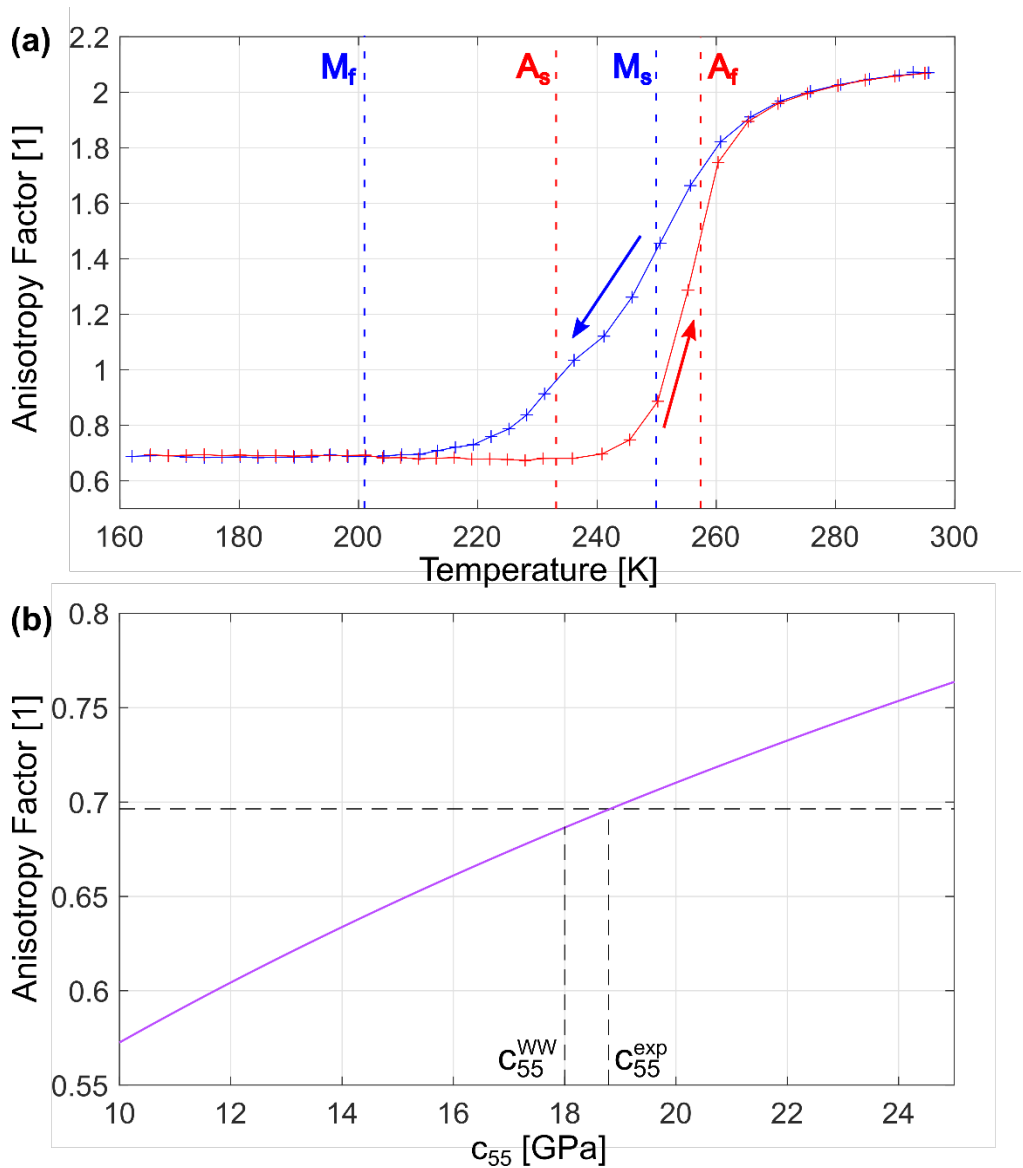

Fig. 4. Temperature evolution of the anisotropy factor *A* (a). Computed evolution of anisotropy factor *A* with the variation of the elastic constant $c_{55}$ (b). The value of *A* at 0 K extrapolated from the experiment is marked with a horizontal dashed line, the corresponding value of $c_{55}$ is denoted by superscript "exp", the DFT value of $c_{55}$ obtained in [12] is denoted by superscript "WW"; see the text for further details.

In addition to changes in elasticity, we also observed changes in internal friction during MT. The internal friction is directly related to the changes in the width of resonance peaks, which allows its evolution to be easily extracted for the RUS data. Fig. 3(c) shows the increase in internal friction during MT as well as when approaching MT from both sides, which is typical for martensitic transitions in all shape memory alloys. The literature has documented a similar evolution for electrical resistance changes during MT (e.g., Fig. 3 in [2]).

In summary, the elasticity in single crystal NiTi during temperature-induced MT evolves as follows: from anisotropic elasticity of austenite with $c_{44} > c'$ ($A \approx 2$), through nearly isotropic elastic behavior of unstable austenite at the beginning of MT to the final switching of soft and hard shear modes in martensite ($A \approx 0.7$). This behavior can also be documented by changes in the velocity curves in the principal plane $(001)_{B2}$ in Fig. 2(b).

Theoretical calculation of elasticity of self-accommodated martensite

The orientation of soft shears in martensite is different compared with austenite. Although the martensitic self-accommodated microstructure is formed by equally populated martensitic variants leading to the preservation of macroscopic cubic symmetry, it still has a relatively large elastic anisotropy.

To reveal the origin of this anisotropy and to relate the obtained experimental data to theoretical *ab-initio* calculations of the elasticity of the B19' structure [12,15], we have used the Hill homogenization scheme to estimate the elasticity of the self-accommodated microstructure from the single-variant elasticity. For this calculation, we used the DFT-based elastic constants of monoclinic B19' martensite from [12], which are listed in the first line in Tab. 2. Note that in this table, the elastic constants are shown for a different orientation of the orthogonal coordinate system than used in [12]. With respect to the commonly used crystallographic system (see, e.g., Fig. 4 in [24]), the orthogonal system in [12] is defined as $\mathbf{x}_1^{WW} \parallel [001]_{B19'}$, $\mathbf{x}_2^{WW} \parallel [010]_{B19'}$, $\mathbf{x}_3^{WW} \parallel$ normal to $(100)_{B19'}$ planes. In Tab. 2, instead, the elastic constants are expressed in a rotated coordinate system, where $\mathbf{x}_1$ is perpendicular to the $(001)_{B19'}$ plane, $\mathbf{x}_2 \parallel [010]_{B19'}$ and $\mathbf{x}_3 \parallel [100]_{B19'}$. In this coordinate system, the instability of the B19' lattice with respect to $(001)[100]_{B19'}$ is directly represented by one of the elastic coefficients, $c_{55}$. For this reason, we will use this coordinate system throughout the whole analysis.

Since the microstructures in temperature-induced NiTi martensites contain fine $(001)_{B19'}$ laminates [25,26], the elasticity of such a laminate was calculated first. The laminate $(001)_{B19'}$ has an orthorhombic symmetry (assuming the same content of both variants forming the laminate), and its elastic tensor is the orthorhombic part of the elastic tensor of single variant B19' (the second row in Tab. 2). The elasticity of the self-accommodated microstructure was then calculated by Hill homogenization (average of the Reuss and Voigt limits) of the elasticity of six $(001)_{B19'}$ equivalent laminates differently oriented with respect to austenite. Their orientation was taken from the lattice correspondence of austenite and martensite ($\mathbf{x}_1 \parallel \{110\}_{B2}$, $\mathbf{x}_2 \parallel \{1\text{-}10\}_{B2}$, $\mathbf{x}_3 \parallel \{001\}_{B2}$), neglecting the possible small rotations of the laminates relative to the austenite due to deformation during MT. The resulting self-accommodated microstructure has cubic anisotropy with constants $c_{11}$ = 233 GPa, $c_{12}$ = 109 GPa, and $c_{44}$ = 42 GPa; see the third line in Tab. 2. For validation purposes, the same calculation procedure was applied for the more recent DFT-based elastic constants of B19' martensite published in [15]. The results are even closer to our experiment: $c_{11}$ = 225 GPa, $c_{12}$ = 117 GPa, and $c_{44}$ = 39 GPa; see the fourth line of Tab. 2.

Tab. 2: Comparison of elastic properties of martensite from the experiment (linear extrapolation to 0K) with the theoretical prediction (DFT + homogenization). Elastic constants from [12] have been rotated to a selected coordinate system; see the text for details. All values are in GPa.

| $c_{ij}$ | $c_{11}$ | $c_{12}$ | $c_{13}$ | $c_{15}$ | $c_{22}$ | $c_{23}$ | $c_{25}$ | $c_{33}$ | $c_{35}$ | $c_{44}$ | $c_{46}$ | $c_{55}$ | $c_{66}$ | $c'$ |
|---|---|---|---|---|---|---|---|---|---|---|---|---|---|---|
| B19' [12] | 235 | 126 | 96 | 14 | 241 | 128 | -9 | 195 | 13 | 77 | -4 | 18 | 76 | - |
| Orthorhombic part of [12] | 235 | 126 | 96 | | 241 | 128 | | 195 | | 77 | | 18 | 76 | |
| B19' Hill avg. [12] | 233 | 109 | | | | | | | | 42 | | | | 62 |
| B19' Hill avg. [15] | 225 | 117 | | | | | | | | 39 | | | | 54 |
| B19' RUS experiments (extrapolation to 0 K) | 229 | 117 | | | | | | | | 39 | | | | 56 |

The last line of Tab. 2 presents linear extrapolation of the experimental results to 0 K. Comparing the experimental measurements and theoretical predictions, we find a pretty good agreement. Clearly, the experiment on the effective elasticity of twined martensite providing three independent elastic constants cannot verify the theoretical computation of elasticity of the monoclinic single variant with 13 independent elastic constants. However, we can focus on the most delicate issue of the DFT calculations of the elasticity of NiTi martensite, which is the instability of the B19' structure. DFT calculations [12-15] consistently show that the B19' structure in NiTi is not the ground state, and the structure exhibits an instability with respect to the $(001)[100]_{B19'}$ shear, that transforms the monoclinic B19' structure into an energetically more stable orthorhombic B33 structure. The importance of this instability recently gained even more attention when the kwinking mechanism was identified as the basic plastic-forming mechanism of NiTi martensite [27], in which $(001)[100]_{B19'}$ slip plays a crucial role. The $(001)[100]_{B19'}$ shear is related to the elastic constant $c_{55}$. Its theoretical predictions vary in the literature; however, $c_{55}$ constant usually [12,15] corresponds to the lowest shear constant in martensite. Fig. 4(b) shows how the variation of $c_{55}$ affects the effective anisotropy of the twinned martensite calculated by the homogenization described above (with other constants fixed). As seen in Fig 4(b), martensite's experimentally determined anisotropy ratio is almost constant down to the minimum investigated temperature. For this reason, we can also extrapolate the value A = 0.7 to 0 K. The theoretical (homogenization) value of A corresponding to this extrapolated experimental anisotropy value gives $c_{55}$ = 18.8 GPa; see Fig. 4 (b). This value is very close to the computational prediction published in [12] (18.0 GPa, see Table 2); thus, we can conclude that the relatively strong elastic anisotropy of the fine martensitic microstructure that we have determined experimentally can be understood as a direct consequence of the soft shear behavior of the B19' lattice represented by the low value of the constant $c_{55}$.

**Conclusions**

We have used ultrasonic methods to characterize the elastic properties of NiTi SMA within a temperature cycle, which induces a reversible transformation between austenite and martensite. We have observed pronounced changes in elasticity leading to a switch in the material anisotropy, verified that the cubic symmetry is not violated by the MT, and discussed the measured values of elastic constants in relation to the recent theoretical findings.

During the cooling of the sample, first, we observed an accelerating decrease of the elastic constant $c_{44}$. This precursor of MT in NiTi occurs as much as 30 K before MT can be detected by conventional techniques such as DSC. After the transformation is initiated, austenite exhibits the lowest shear moduli, and its elasticity is nearly isotropic; in the final stage of the transformation, the constant increases again. This is in sharp contrast to other SMAs (e.g., NiMnGa, CuAlNi), where the shear instability of austenite before the MT usually leads to a dramatic increase in elastic anisotropy.

A similar precursor effect is observed when martensite is heated. The $c_{44}$ constant declines from the linear trend even before the reverse transformation is initiated, the decrease gradually accelerates, and the decline turns into a rise even before full austenite is reached. Such a strikingly non-monotonous character is not observed for other elastic constants since the dominant shear mechanism of MT involves volume-preserving stretching in the $[111]_{B2}$ direction, which manifests itself dominantly in $c_{44}$.

Cooling of the free-standing sample induces self-accommodated martensite, which is generally (in most directions) stiffer than austenite – this agrees well with the theoretical assumption of the higher stability of the martensitic phase. Both the shear and longitudinal elastic constants increase with decreasing temperature. The increase in $c'$ in martensite during MT corresponds to the stabilization of the shear instability in austenite; however, $c_{44}$ reaches similar values in both austenite and martensite. This leads to a switch in the material anisotropy, i.e., a qualitative change in the character of the elastic anisotropy of the material: whereas $A \approx 2$ in austenite, $A \approx 0.7$ in martensite.

The self-accommodated martensitic microstructure retains cubic symmetry (within the experimental accuracy). This indicates equal populations of different martensitic variants in the measured martensitic microstructure. However, although no variant appears to be favored, the overall elastic anisotropy of the self-accommodated martensitic microstructure is still relatively strong after MT, and it remains approximately constant even during cooling well below the transformation region.

Using a simplified model of self-accommodated microstructure and Hill's homogenization scheme, we were able to compare the results of theoretical analysis of martensite elasticity based on DFT calculations with our experimental data. Such a comparison provides the following insights:

1) The lower value of $c_{44}$ compared to $c'$ in self-accommodated martensite is due to the characteristic elastic anisotropy of the B19' NiTi crystal structure. The B19' structure exhibits instability with respect to the $(001)[100]_{B19'}$ shear resulting in a quite low value of the shear modulus $c_{55}$. This low shear modulus contributes significantly to the macroscopic constant $c_{44}$ of the self-accommodated martensite and its experimentally observed elastic anisotropy.

2) The experimental values of martensitic elastic constants of the self-accommodated microstructure extrapolated to 0 K agree well with the theoretical predictions obtained by homogenization of the single-variant elasticity calculated for the B19' structure with DFT [12,15]. Indeed, the experimental values are very close to Hill's mean. Such a high plausibility of DFT predictions of the martensitic elasticity is by far not common; see, e.g., the recent work on 10M NiMnGa martensite [28] and the discussion there. This success of the theoretical prediction is even more remarkable given the main challenges it faces: B19' structure is not the ground state in the DFT calculations and shows instability with respect to the $(001)[100]_{B19'}$ shear.

**Acknowledgment:** This work has been financially supported by the Czech Science Foundation [project No. 22-20181S] and by the Operational Programme Johannes Amos Comenius of the Ministry of Education, Youth and Sport of the Czech Republic, within the frame of project Ferroic Multifunctionalities (FerrMion) [project No. CZ.02.01.01/00/22_008/0004591], co-funded by the European Union. The SEM measurements were supported by CzechNanoLab Research Infrastructure MEYS CR (LM2023051).

# References

1. Mercier O, Melton KN, Gremaud G, Hägi J (1980) Single-crystal elastic constants of the equiatomic NiTi alloy near the martensitic transformation. Journal of Applied Physics 51: 1833 – 1834. DOI: 10.1063/1.327750.
2. Brill TM, Mittelbach S, Assmus W, Mullner M, Luthi B (1991) Elastic properties of NiTi. Journal of Physics: Condensed Matter 3: 9621 – 9627. DOI: 10.1088/0953-8984/3/48/004.
3. Ren X, Otsuka K (1998) The role of softening in elastic constant c44 in martensitic transformation. Scripta Materialia 38: 1669 – 1675. DOI: 10.1016/S1359-6462(98)00078-5.
4. Planes A, Mañosa L, Ríos-Jara D, Ortín J (1992) Martensitic transformation of Cu-based shape-memory alloys: Elastic anisotropy and entropy change. Physical Review B 45: 7633 – 7639. DOI: 10.1103/PhysRevB.45.7633.
5. Chernenko VA, Pons J, Seguí C, Cesari E (2002) Premartensitic phenomena and other phase transformations in Ni-Mn-Ga alloys studied by dynamical mechanical analysis and electron diffraction. Acta Materialia 50: 53 – 60. DOI: 10.1016/S1359-6454(01)00320-2.
6. Seiner H, Sedlák P, Bodnárová L, Drahokoupil J, Kopecký V, Kopeček J, Landa M, Heczko O (2013) The effect of antiphase boundaries on the elastic properties of Ni-Mn-Ga austenite and premartensite. Journal of Physics Condensed Matter 25: 425402. DOI: 10.1088/0953-8984/25/42/425402.
7. Stebner AP, Brown DW, Brinson LC (2013) Measurement of elastic constants of monoclinic nickel-titanium and validation of first-principles calculations. Applied Physics Letters 102: 211908. DOI: 10.1063/1.4808040.
8. Šittner P, Heller L, Pilch J, Curfs C, Alonso T, Favier D (2014) Young's modulus of austenite and martensite phases in superelastic NiTi wires. Journal of Materials Engineering and Performance 23: 2303 – 2314. DOI: 10.1007/s11665-014-0976-x.
9. Bucsek AN, Paranjape HM, Stebner AP (2016) Myths and Truths of Nitinol Mechanics: Elasticity and Tension–Compression Asymmetry. Shape Memory and Superelasticity 2: 264 – 271. DOI: 10.1007/s40830-016-0074-z.
10. Thomasová M, Seiner H, Sedlák P, Frost M, Ševčík M, Szurman I, Kocich R, Drahokoupil J, Šittner P, Landa M (2017) Evolution of macroscopic elastic moduli of martensitic polycrystalline NiTi and NiTiCu shape memory alloys with pseudoplastic straining. Acta Materialia 123: 146 – 156. DOI: 10.1016/j.actamat.2016.10.024.
11. Grabec T, Sedlák P, Zoubková K, Ševčík M, Janovská M, Stoklasová P, Seiner H (2021) Evolution of elastic constants of the NiTi shape memory alloy during a stress-induced martensitic transformation. Acta Materialia 208: 116718. DOI: 10.1016/j.actamat.2021.116718
12. Wagner MF-X, Windl W (2008) Lattice stability, elastic constants and macroscopic moduli of NiTi martensites from first principles. Acta Materialia 56: 6232 – 6245. DOI: 10.1016/j.actamat.2008.08.043.
13. Hatcher N, Kontsevoi OY, Freeman AJ (2009) Role of elastic and shear stabilities in the martensitic transformation path of NiTi. Physical Review B - Condensed Matter and Materials Physics 80: 144203. DOI: 10.1103/PhysRevB.80.144203.
14. Wang J, Sehitoglu H (2014) Martensite modulus dilemma in monoclinic NiTi-theory and experiments. International Journal of Plasticity 61: 17 – 31. DOI: 10.1016/j.ijplas.2014.05.005.
15. Haskins JB, Lawson JW (2017) Finite temperature properties of NiTi from first principles simulations: Structure, mechanics, and thermodynamics. Journal of Applied Physics 121: 205103. DOI: 10.1063/1.4983818.
16. Auricchio F, Sacco E (1997) A one-dimensional model for superelastic shape-memory alloys with different elastic properties between austenite and martensite. International Journal of Non-Linear Mechanics 32: 1101-1114. DOI: 10.1016/S0020-7462(96)00130-8.
17. Frost M, Sedlák P, Kadeřávek L, Heller L, Šittner P (2016) Modeling of mechanical response of NiTi shape memory alloy subjected to combined thermal and non-proportional mechanical loading: A case study on helical spring actuator. Journal of Intelligent Material Systems and Structures 27: 1927 – 1938. DOI: 10.1177/1045389X15610908

18. Šittner P, Sedlák P, Seiner H, Sedmák P, Pilch J, Delville R, Heller L, Kadeřávek L (2018) On the coupling between martensitic transformation and plasticity in NiTi: Experiments and continuum based modeling. Progress in Materials Science 98: 249 – 298. DOI: 10.1016/j.pmatsci.2018.07.003.
19. Heller L, Šittner P, Sedlák P, Seiner H, Tyc O, Kadeřávek L, Sedmák P, Vronka M (2019) Beyond the strain recoverability of martensitic transformation in NiTi. International Journal of Plasticity 116: 232 – 264. DOI: 10.1016/j.ijplas.2019.01.007.
20. Sedlák P, Seiner H, Zídek J, Janovská M, Landa M (2014) Determination of All 21 Independent Elastic Coefficients of Generally Anisotropic Solids by Resonant Ultrasound Spectroscopy: Benchmark Examples. Experimental Mechanics 54: 1073 – 1085. DOI: 10.1007/s11340-014-9862-6.
21. Sedlák P., Janovská M., Bodnárová L., Heczko O., Seiner H (2020) Softening of Shear Elastic Coefficients in Shape Memory Alloys Near the Martensitic Transition: A Study by Laser-Based Resonant Ultrasound Spectroscopy. Metals 10: 1383. DOI: 10.3390/met10101383.
22. Grabec T, Soudná Z, Repček K, Lünser K, Fähler S, Stoklasová P, Sedlák P, Seiner H (2024) Guided acoustic waves in thin epitaxial films: Experiment and inverse problem solution for NiTi. Ultrasonics 138: 107211. DOI: 10.1016/j.ultras.2023.107211.
23. Sedlák P, Seiner H, Landa M, Novák V, Šittner P, Mañosa Ll (2005) Elastic constants of bcc austenite and 2H orthorhombic martensite in CuAlNi shape memory alloy. Acta Materialia 53: 3643 – 3661. DOI: 10.1016/j.actamat.2005.04.013.
24. Ko W-S, Grabowski B, Neugebauer (2015) Development and application of a Ni-Ti interatomic potential with high predictive accuracy of the martensitic phase transition. Physical Review B - Condensed Matter and Materials Physics 92: 134107. DOI: 0.1103/PhysRevB.92.134107.
25. Xie Z, Liu Y, Van Humbeeck J (1998) Microstructure of NiTi shape memory alloy due to tension-compression cyclic deformation. Acta Materialia 46: 1989 – 2000. DOI: 10.1016/S1359-6454(97)00379-0.
26. Krishnan M, Singh JB (2000) A novel B19' martensite in nickel titanium shape memory alloys. Acta Materialia 48: 1325 – 1344. DOI: 10.1016/S1359-6454(99)00423-1.
27. Seiner H, Sedlák P, Frost M, Šittner P (2023) Kwinking as the plastic forming mechanism of B19' NiTi martensite. International Journal of Plasticity 168: 103697. DOI: 10.1016/j.ijplas.2023.103697.
28. Repček K, Stoklasová P, Grabec T, Sedlák P, Olejňák J, Vinogradova M, Sozinov A, Veřtát P, Straka L, Heczko O, Seiner H (2024) Compliant Lattice Modulations Enable Anomalous Elasticity in Ni–Mn–Ga Martensite. Advanced Materials 36: 2406672. DOI: 10.1002/adma.202406672.